\documentclass[prl,twocolumn,superscriptaddress]{revtex4}
\usepackage{amsmath,amssymb,graphicx,xcolor,bm,soul}
\newcommand{\ah}[1]{\hat{a}^{\vphantom{\dag}}_{#1}}
\newcommand{\ahh}[1]{\hat{a}^{2}_{#1}}
\newcommand{\ad}[1]{\hat{a}^\dag_{#1}}
\newcommand{\add}[1]{\hat{a}^{\dag2}_{#1}}
\newcommand{\bh}[1]{\hat{b}^{\vphantom{\dag}}_{#1}}
\newcommand{\bhh}[1]{\hat{b}^{2}_{#1}}
\newcommand{\bd}[1]{\hat{b}^\dag_{#1}}

\begin{document}

\title{Quantum Exciton-Polariton Networks through Inverse Four-Wave Mixing}

\author{T. C. H. Liew}
\affiliation{Division of Physics and Applied Physics, School of Physical and Mathematical Sciences, Nanyang Technological University, 21 Nanyang Link, Singapore 637371}

\author{Y. G. Rubo}
\affiliation{Center for Theoretical Physics of Complex Systems, Institute for Basic Science (IBS), Daejeon 34051, Republic of Korea}
\affiliation{Instituto de Energ\'{\i}as Renovables, Universidad Nacional Aut\'onoma de M\'exico, Temixco, Morelos 62580, Mexico}

\begin{abstract}
We demonstrate the potential of quantum operation using lattices of exciton-polaritons in patterned semiconductor microcavities. By introducing an inverse four-wave mixing scheme acting on localized modes, we show that it is possible to develop non-classical correlations between individual condensates. This allows a concept of quantum exciton-polariton networks, characterized by the appearance of multimode entanglement even in the presence of realistic levels of dissipation.
\end{abstract}

\date{\today}

\maketitle

Recently, there has been a significant attention devoted to the study of exciton-polaritons in lattices~\cite{Lai2007,CerdaMendez2010,Kim2011,Kim2013,CerdaMendez2013,Jacqmin2014,Winkler2016}. As systems of nonlinear interacting bosons, they have often been suggested as potential candidates of quantum simulators~\cite{Feynmann1982,Lloyd1996} and indeed the minimization of the energy of a particular Hamiltonian on a graph was a problem considered recently~\cite{Berloff2016}. While the majority of studies of exciton-polaritons have been restricted to the classical regime~\cite{Utsunomiya2011,Marandi2014}, the quantum nature of polaritons has received revived attention recently~\cite{Cuevas2016}. Therefore, it is natural to question whether exciton-polaritons can be used to form lattices of entangled modes. Here we must be aware that a lattice or graph of polaritons does not behave as a system of qubits. Instead each node of a polariton network could be described by the quantum field amplitude $\hat{a}_n$ or the continuous amplitude and phase variables associated with the operators:
\begin{equation}
\hat{q}_n=\frac{\hat{a}_n+\hat{a}_n^\dagger}{\sqrt{2}}, \qquad
\hat{p}_n=\frac{\hat{a}_n-\hat{a}_n^\dagger}{i\sqrt{2}}.
\end{equation}
Since continuous variable modes can be entangled, networks of continuous variable modes are highly relevant for quantum applications. As an example, cluster state computation~\cite{Raussendorf2001} based on continuous variables~\cite{Menicucci2006} is a potential route towards universal computation. It relies on producing a highly entangled state from an arbitrary lattice or graph of modes coupled by two-mode squeezing type interactions, with a Hamiltonian of the form:
\begin{equation}
\mathcal{H_S}=\sum_{nm}w_{nm}\left(\hat{a}_n\hat{a}_m+\hat{a}^\dagger_n\hat{a}^\dagger_m\right),
\end{equation}
where $w_{nm}$ describes the weights of different connections in the graph. Arranging such a Hamiltonian is already a problem and it must be done making use of some interaction process that is stronger than any detrimental processes in the system (dissipation, dephasing, etc.). While evidence of strongly interacting polaritons~\cite{Sun2015} was reported recently, it is not clear if any nonlinear interaction process in microcavities is sufficiently strong for the generation of quantum resources. In the absence of strong interactions, exciton-polaritons tend to only demonstrate nonlinear effects at high densities, when they are well described by the classical physics corresponding to the mean-field approximation. For this reason only a handful of experimental reports of quantum exciton-polariton effects have appeared in the literature~\cite{Karr2004b,Savasta2005}.

\emph{Two-Mode Squeezing.---}Before considering how to build a polariton network, it is instructive to consider the effect of the two-mode squeezing type Hamiltonian:
\begin{equation}
 \hat{H}=-\frac{i\alpha}{2}\big(\ah{1}\ah{2}-\ad{1}\ad{2}\big).\label{eq:Ham2Mode}
\end{equation}
Such a Hamiltonian generates entanglement, which can be characterized by the violation of the inequality~\cite{Duan2000,Simon2000}
\begin{equation}\label{eq:inequalityS}
 1\leq S_{12}=\frac{1}{2}\left[V(\hat{q}_1-\hat{q}_2)+V(\hat{p}_1+\hat{p}_2)\right],
\end{equation}
where the variances are defined by $V(\hat{O})=\langle\hat{O}^2\rangle-\langle\hat{O}\rangle^2$.

The Heisenberg equations of motion give the evolution of the quantum field operators $\ah{1,2}(t)\equiv e^{i\hat{H}t}\ah{1,2}e^{-i\hat{H}t}$ (we set $\hbar=1$)
\begin{equation}\label{eq:fieldevol}
\ah{1,2}(t)=\cosh(\alpha t/2)\ah{1,2}+\sinh(\alpha t/2)\ad{2,1}.
\end{equation}

To calculate the second order correlators, we consider operators $\hat{K}=1+\ad{1}\ah{1}+\ad{2}\ah{2}$, $\hat{L}=\ah{1}\ah{2}+\ad{1}\ad{2}$, $\hat{M}=i(\ah{1}\ah{2}-\ad{1}\ad{2})$ and use the Lie algebra $[\hat{M},\hat{K}]=2i\hat{L}$, $[\hat{M},\hat{L}]=2i\hat{K}$ to obtain
\begin{subequations}\label{eq:KLevol}
\begin{align}
  \hat{K}(t) &= \cosh(\alpha t)\hat{K} + \sinh(\alpha t)\hat{L}, \\
  \hat{L}(t) &= \cosh(\alpha t)\hat{L} + \sinh(\alpha t)\hat{K}.
\end{align}
\end{subequations}
Using these relations and taking the vacuum state as an initial condition one arrives at
\begin{equation}
  S_{12}=\big<\hat{K}(t)-\hat{L}(t)\big>=e^{-\alpha t}<1,
\end{equation}
indicating evolution of the system towards the entangled co-eigenstate of the EPR pair of operators $\hat{q}_1-\hat{q}_2$ and $\hat{p}_1+\hat{p}_2$. This result, $S_{12}=e^{-\alpha t}$, remains unchanged for initial coherent states of the fields. We note that the EPR pair of operators~\cite{Einstein1937} needed to demostrate the nonclassical correlations depends on the Hamiltonian. Other possible pairs can be obtained with the gauge transformation of operators in \eqref{eq:inequalityS}, $\ah{1,2}\rightarrow\ah{1,2}e^{i\phi_{1,2}}$, with subsequent optimization over the phases $\phi_{1,2}$.

\emph{Inverse Four-Wave Mixing.}---Let us now consider a single cavity with a four-wave mixing (parametric) type resonance, described with the Hamiltonian
\begin{equation}
  \mathcal{H}_0=\frac{\alpha_0}{2}\left(\hat{a}^\dagger\hat{a}^\dagger\hat{a}_L\hat{a}_U+\hat{a}_L^\dagger\hat{a}_U^\dagger\hat{a}\hat{a}\right),\label{eq:H0}
\end{equation}
where $\alpha_0$ describes the strength of the four-wave mixing process. Physical realizations of the above Hamiltonian could be made in exciton-polariton micropillars~\cite{Vasconcellos2011} or a Kerr nonlinear photonic crystal cavity~\cite{Gerace2009}. Finding a parametric resonance would however require careful tuning~\cite{Diederichs2006}, which suggests that systems compatible with post-growth tuning would be the most realistic choices. For example, dipolariton based setups allow electrical control of mode energies~\cite{Cristofolini2012}. Regardless of the mechanism of introducing Hamiltonian~\eqref{eq:H0}, it is typically the case that $\alpha_0$ will be weak compared to the system losses $\Gamma$, that is, typical optical systems are only weakly nonlinear ($\alpha_0\ll\Gamma$).

\begin{figure}[h!]
\includegraphics[width=\columnwidth]{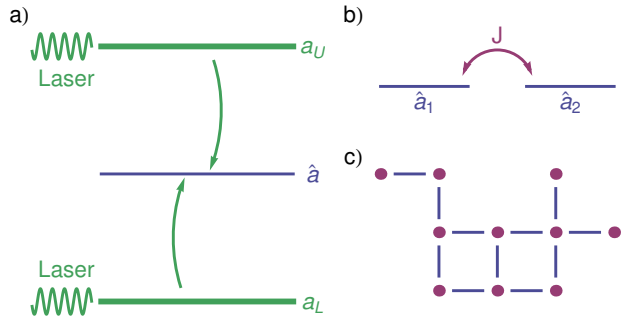}
\caption{(color online) a) Scheme of inverse four-wave mixing. b) Coupling of spatially separated modes, each driven by the inverse four-wave mixing scheme of (a). c) Potential generalization into lattices and arbitrary graphs.}
\label{fig:scheme}
\end{figure}

The Hamiltonian~\eqref{eq:H0} is usually considered for generating the fields $\hat{a}_L$ and $\hat{a}_U$ from initial excitation of the field $\hat{a}$, however, we can also consider the inverse process illustrated in Fig.~\ref{fig:scheme}a.
Namely, if the modes $\hat{a}_L$ and $\hat{a}_U$ are driven by coherent laser fields then particles scatter in pairs from $\hat{a}_L$ and $\hat{a}_U$ to the mode $\hat{a}$. It is true that under such conditions the modes $\hat{a}_L$ and $\hat{a}_U$ should behave only classically, such that their physics can not go beyond what is expected from making the mean-field approximation on these modes, but doing so leaves still a reduced quantum Hamiltonian acting on the mode $\hat{a}$:
\begin{equation}
  \mathcal{H}_0=\frac{\alpha}{2}\left(\hat{a}^\dagger\hat{a}^\dagger+\hat{a}\hat{a}\right),
  \label{eq:Ham2photonpump}
\end{equation}
where $\alpha=\alpha_0\langle a_U\rangle\langle a_L\rangle$. While this is just the Hamiltonian of two particle creation, by considering its introduction via the aforementioned inverse four-wave mixing process we have a way to make this a strong effect: since $\langle a_L\rangle$ and $\langle a_U\rangle$ can be increased by the resonant driving intensity, one can reach the regime $\alpha\gg\Gamma$.

Considering exciton-polariton systems the regime $\alpha\gg\Gamma$ has essentially been realized previously under different conditions, where the blueshift due to polariton-polariton interactions may exceed the linewidth and cause bistability~\cite{Baas2004b}. It is worth mentioning that four-wave mixing experiments also revealed an interesting polarization dependence~\cite{Krizhanovskii2006,Leyder2007}, which allow the signal mode $\hat{a}$ to have a different linear polarization to that of the others ($\hat{a}_U$ and $\hat{a}_L$), useful for better resolution and limiting other scattering processes.

\emph{Coupled cavities.}---If we now consider a pair of coupled cavities, which could be made with the techniques of Ref.~\cite{Vasconcellos2011}, the model Hamiltonian becomes (see Fig.~\ref{fig:scheme}b):
\begin{equation}\label{eq:H}
 \mathcal{H}=\frac{\alpha}{2}\left(\ahh{1}+\ahh{2}+\add{1}+\add{2}\right)
 -J\left(\hat{a}_1^\dagger\hat{a}_2+\hat{a}_2^\dagger\hat{a}_1\right),
\end{equation}
where $J$ is the coupling constant between the cavities and we can set $\alpha>0$ without loss of generality. We show below that this Hamiltonian results in entanglement between modes $\hat{a}_1$ and $\hat{a}_2$.

It is convenient to define new operators, representing a symmetric-antisymmetric basis
\begin{equation}
 \hat{a}_1=\frac{\hat{a}_+ + \hat{a}_-}{\sqrt{2}}, \qquad
 \hat{a}_2=\frac{\hat{a}_+ - \hat{a}_-}{\sqrt{2}},
\end{equation}
decoupling the Hamiltonian into two parts:
\begin{equation}
 \mathcal{\hat{H}}=\frac{1}{2}\sum_{\sigma=\pm}\left[ \alpha(\ahh{\sigma}+\add{\sigma}) - 2\sigma J\ad{\sigma}\ah{\sigma} \right].
\end{equation}
We can then consider the Bogoliubov transform
\begin{equation}
 \hat{a}_\sigma=\cosh(x/2)\hat{b}_\sigma+\sigma\sinh(x/2)\hat{b}^\dagger_\sigma,
\end{equation}
which in the case $|J|>\alpha$ and $\tanh(x)=\alpha/J$ reduces the Hamiltonian into the simple form
\begin{equation}\label{eq:HamBogoliubov}
 \mathcal{H}=\omega\big(\bd{+}\bh{+}-\bd{-}\bh{-}\big),
\end{equation}
where $\omega=\sqrt{J^2-\alpha^2}$.

If we take the vacuum state as the initial condition then only the second-order correlators of $a$-fields contribute to the inequality \eqref{eq:inequalityS}. It is easy to show that
%
%
%
\begin{subequations}
\begin{align}
 \langle \bd{\sigma}\bh{\sigma}(t)\rangle &= \sinh^2(x/2),  \\
 \langle \bhh{\sigma}(t)\rangle &= -\frac{\sigma}{2}\sinh(x)e^{-2i\sigma\omega t},
\end{align}
\end{subequations}
which results in
\begin{subequations}
\begin{align}
 & \langle\ad{\sigma}\ah{\sigma}(t)\rangle=\sinh^2(x)\sin^2(\omega t), \\
 & \langle\ahh{\sigma}(t)\rangle=\sinh(x)\sin(\omega t)\left[\sigma\cosh(x)\sin(\omega t)+i\cos(\omega t)\right].
\end{align}
\end{subequations}
In the symmetric-antisymmetric basis we have
\begin{equation}\label{eq:S12aplusaminus}
 S_{12}=1+\langle\ad{+}\ah{+}+\ad{-}\ah{-}-\Re\big(\ahh{+}-\ahh{-}\big)\rangle,
\end{equation}
where $\Re$ denotes the real part and the first order correlators vanish in our case. Substituting the explicit form of the correlators, we obtain:
\begin{align}
 S_{12}&=1+2\sinh^2(x)\sin^2(\omega t)-\sinh(2x)\sin^2(\omega t) \nonumber \\
       &=\frac{J+\alpha\cos(2\omega t)}{J+\alpha}.
\end{align}

\begin{figure}[h!]
\includegraphics[width=\columnwidth]{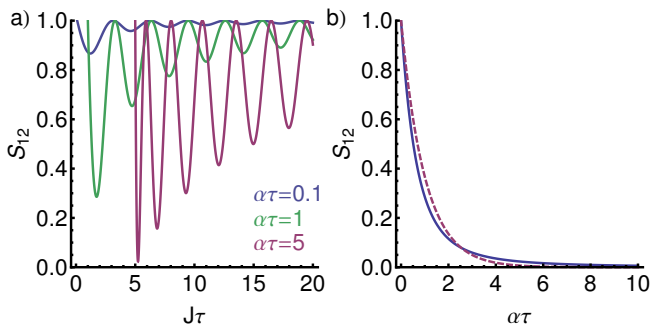}
\caption{(color online) a) Dependence of $S_{12}$ on $J$. Here $S_{12}$ is evaluated at $t=\tau$ and it can be seen that there is an optimal value of $J\approx\alpha$. The minimum value of $S_{12}$ can become smaller for increasing values of $\alpha$. b) Dependence of the optimum value of $S_{12}$ (with optimally chosen $J$) on $\alpha$. The dashed curve shows the function $S_{12}=e^{-\alpha \tau}$ obtained from the ideal squeezing Hamiltonian (\ref{eq:Ham2Mode}) for comparison.}
\label{fig:S12}
\end{figure}

While this expression can never reach the value of zero, for the case $J>\alpha$, one can reach the value $(J-\alpha)/(J+\alpha)$ for the specific time when the cosine function evaluates to $-1$. Since $J-\alpha$ can be tuned to be small, one can then in principle reach arbitrarily small values of $S_{12}$. To give a visualization, Fig.~\ref{fig:S12}a shows the variation in $S_{12}$ at some fixed time as a function of $J$. Figure~\ref{fig:S12}b then shows the minimal value of $S_{12}$ obtainable for increasing values of $\alpha$.

The formation of entanglement in the above scheme might be seen as a round about way to create entanglement from four-wave mixing, which could be obtained already from Hamiltonian~\eqref{eq:H0}. Indeed the usual method of exciting the central mode $\hat{a}$ and looking at correlations between $\hat{a}_L$ and $\hat{a}_U$ has been considered before, in different contexts~\cite{Savasta1997,Schwendimann2003,Karr2004,Portolan2009,Romanelli2010}. It should be stressed however that the conventional method requires $\alpha_0$ to be significant compared to the dissipation rate and also $\alpha_0$ should be stronger than other scattering processes (e.g., scattering with acoustic phonons) that may resonantly couple the modes to be entangled. In the scheme that we consider here $\alpha$ can become the dominant interaction in the system as it is enhanced by the density of modes $a_L$ and $a_U$. Furthermore, local interactions, such as scattering with phonons and sample disorder are not able to couple spatially separated modes $\hat{a}_1$ and $\hat{a}_2$.

\emph{Dissipation.}---We have shown so far that the system of coupled cavities driven by parametric resonance can generate entangled states, which become asymptotically close to the level of entanglement expected from a two-mode squeezing type operation, as measured by the violation of inequality~\eqref{eq:inequalityS}. As we have noted in the previous section $\alpha$ can be controlled by the intensity of external lasers. In principle, $J$ can also be controlled by external fields, for example, by using external electric~\cite{Christmann2010} or optical fields~\cite{Amo2010} to modify the potential between lattice points.

\begin{figure}[h!]
\includegraphics[width=\columnwidth]{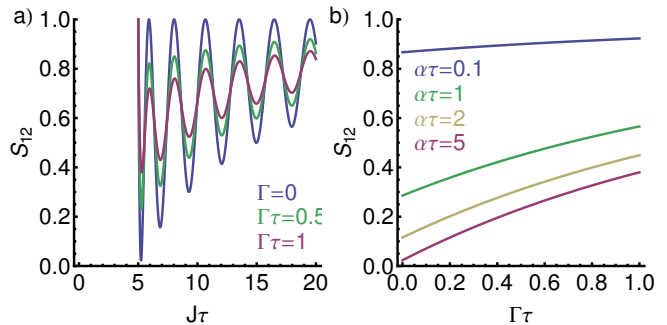}
\caption{(color online) a) Dependence of $S_{12}$ on $J$. As in Fig.~\ref{fig:S12}a $S_{12}$ is evaluated at $t=\tau$, but here we consider a fixed value of $\alpha\tau=5$ and consider different values of dissipation $\Gamma$. b) Dependence of the optimum value of $S_{12}$ (with optimally chosen $J$) on $\Gamma$.}
\label{fig:S12Gamma}
\end{figure}

While we expect the regime $\alpha\gg\Gamma$ to be experimentally accessible, given the parametric driving scheme, it is still instructive to consider the influence of dissipation in the system. This is readily introduced by modification of the Heisenberg equations:
\begin{equation}\label{eq:HeisenbergDissipation}
 \frac{d\langle\hat{O}\rangle}{dt}=i\langle\big[\hat{\mathcal{H}},\hat{O}\big]\rangle
 +\frac{\Gamma}{2}\sum_n\langle2\ad{n}\hat{O}\ah{n}-\ad{n}\ah{n}\hat{O}-\hat{O}\ad{n}\ah{n}\rangle.
\end{equation}
This introduces additional dissipation terms in our equations of motion, which are solved in the Supplementary Material~\cite{SupplementaryMaterial}. The resulting effect of dissipation is illustrated in Figs.~\ref{fig:S12Gamma}(a,b).
As one would expect, too much dissipation results in a loss of entanglement. However, given the parametric pumping scheme it is in principle possible to work in the limit where $\Gamma \ll \alpha$. At some short time such that $\tau \ll 1/\Gamma$ one then obtains a high degree of entanglement despite the presence of dissipation.

\emph{Multimode Entanglement.}---In comparison to conventional methods of entanglement generation with respect to four-wave mixing, a further advantage of our scheme that entangles polariton modes separated in real space is that it is in principle scalable; by coupling more cavities in space, arbitrary networks could be considered such as the one illustrated in Fig.~\ref{fig:scheme}c.

As an example let us consider a system of four identical cavity, which are subjected to the Hamiltonian:
\begin{align}
\mathcal{H}_4&=\sum_{n=1}^4\frac{\alpha}{2}\left(\hat{a}_n^\dagger\hat{a}_n^\dagger+\hat{a}_n\hat{a}_n\right)\notag\\
&\hspace{5mm}-J_A(t)\left(\hat{a}_1^\dagger\hat{a}_2+\hat{a}_2^\dagger\hat{a}_1+\hat{a}_3^\dagger\hat{a}_4+\hat{a}_4^\dagger\hat{a}_3\right)\notag\\
&\hspace{5mm}-J_B(t)\left(\hat{a}_1^\dagger\hat{a}_3+\hat{a}_3^\dagger\hat{a}_1+\hat{a}_2^\dagger\hat{a}_4+\hat{a}_4^\dagger\hat{a}_2\right).\label{eq:H4}
\end{align}
This Hamiltonian is a generalization of Hamiltonian~\eqref{eq:H}, where we assume that it is possible to control the linear coupling in time. For simplicity, we will consider ($J_A(t)=J$, $J_B(t)=0$) for the time $0<t<\tau$ and ($J_A(t)=0$, $J_B(t)=J$) for time $\tau<t<2\tau$. It is possible to write Heisenberg equations of motion and their time dependent solution can be obtained analytically~\cite{SupplementaryMaterial}. Alternatively, in the absence of dissipation, it is more efficient to solve for the operator evolution in the Heisenberg picture~\cite{SupplementaryMaterial}.

\begin{figure}[h!]
\includegraphics[width=\columnwidth]{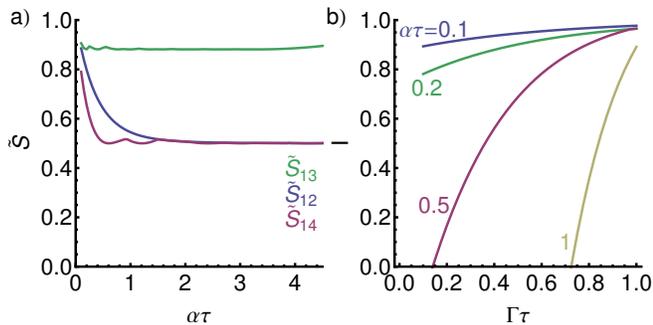}
\caption{(color online) a) Dependence of the optimum values of $\tilde{S}_{12}$, $\tilde{S}_{13}$, and $\tilde{S}_{14}$ (with optimally chosen $J$) on $\alpha$. Parameters: $\Gamma=0$, $t_A=t_B=\tau$. b) Dependence of the optimum value of $I$ (with optimally chosen $J$) on $\Gamma$, for different values of $\alpha$.}
\label{fig:I}
\end{figure}

While violation of inequality~\eqref{eq:inequalityS} is a sufficient condition for entanglement, the definition given of $S_{12}$ is not ideal for all states. In particular, varying the phases of modes $\hat{a}_1$ and $\hat{a}_2$ changes the value of $S_{12}$ and thus to demonstrate the entanglement we should minimize $S_{12}$ over all choices of local phases. As we mentioned above, this is equivalent to finding the best EPR pair of operators. The procedure is detailed in~\cite{SupplementaryMaterial}, where we define $\tilde{S}_{12}$ as the value of $S_{12}$ minimized over phase rotations. The result is shown in Fig.~\ref{fig:I}a, where, in addition to characterizing the entanglement between modes $\hat{a}_1$ and $\hat{a}_2$, we find also entanglement between other pairs of modes, using similar definitions for $S_{13}$ and $S_{14}$ (other combinations of modes display identical entanglement characteristics due to symmetry).

In addition to entanglement between pairs of modes, multimode entanglement, simultaneously between all four modes of the system can be evidenced by the violation of the inequality~\cite{Midgley2010}:
\begin{equation}
\frac{1}{2}\left(V(\hat{q}_1-\hat{q}_2)+V(\hat{p}_1+\hat{p}_2+g\hat{p}_3+g\hat{p}_4)\right)=I\geq1,\label{eq:I}
\end{equation}
where $g$ is an arbitrary real parameter that should be chosen so as to optimize the violation of the inequality. In the general case of four modes, one should also break two other inqualitites to evidence an entangled state, obtained by permuting the modes~\cite{Midgley2010}. However, given the symmetry of our four mode example in a ring these inequalities are equivalent and the violation of inequality~\ref{eq:I} is a sufficient condition. Following correct choice of the parameter $g$ and optimization over the phases of the modes~\cite{SupplementaryMaterial}, we indeed find that the quantity $I$ can drop below one and even reach zero, as shown in Fig.~\ref{fig:I} for different values of $\alpha$ and $\Gamma$.

\emph{Conclusion.}---The evolution of polariton networks from the classical to quantum regime implies finding a mechanism of generating quantum correlations that can overcome the dissipation of the system. Nonlinearity, in the form of polariton-polariton interactions is traditionally weak, however, here we have shown theoretically that an inverse four-wave mixing geometry allows enhancement to an effective strongly nonlinear regime. Local nonlinearity and standard Josephson coupling between spatially separated modes is then sufficient to generate quantum entanglement both between pairs of modes and multiple modes. We hope this will stimulate further discussions on polariton simulators, which have begun recently~\cite{Berloff2016}.

This research was supported by the MOE (Singapore) grant 2015-T2-1-055, by IBS-R024-D1, and by CONACYT (Mexico) grant 251808.


\end{document}